\documentclass[
aps,
prb,
twocolumn,
superscriptaddress,
]{revtex4}

\usepackage{graphicx}
\usepackage{dcolumn}
\usepackage{amsmath}
\usepackage{bm}
\usepackage{times}
\usepackage[varg]{txfonts}

\def\eac{\epsilon_{\mbox{{\scriptsize ac}}}}
\def\xt{\epsilon_{\mbox{{\scriptsize ac}}}^{\mbox{{\scriptsize 2+}}}}
\def\vsp{\vspace{-0.15 in}}

\newcommand{\be}{\begin{equation}}
\newcommand{\ee}{\end{equation}}
\newcommand{\bea}{\begin{eqnarray}}
\newcommand{\eea}{\end{eqnarray}}

\newcommand{\wc}{\omega_{\rm c}}

\newcommand{\Pmw}{{\cal P}_{\omega}}

\def\Edc{\mathcal{E}_{\mathrm{dc}}}

\def\xt{\mathcal{X}_{2}}
\def\eac{\epsilon_{\mathrm{ac}}}
\def\edc{\epsilon_{\mathrm{dc}}}

\def\oc{\omega_{\mbox{\scriptsize {c}}}}

\def\rc{R_{\mbox{\scriptsize {c}}}}

\def\tpi{\tau_{\pi}}

\def\tq{\tau_{\mbox{\scriptsize {q}}}}

\def\ttr{\tau_{\mbox{\scriptsize {tr}}}}

\def\ttr{\tau_{\mbox{\scriptsize {tr}}}}

\newcommand{\req}[1]{Eq.\,(\ref{#1})}

\newcommand{\rfig}[1]{Fig.\,\ref{#1}}

\begin{document}
\title{
Nonlinear response of microwave-irradiated two-dimensional electron systems near the second harmonic of the cyclotron resonance
}
\author{A.\,T. Hatke}
\affiliation{School of Physics and Astronomy, University of Minnesota, Minneapolis, Minnesota 55455, USA}
\author{M.\,A. Zudov}
\email[Corresponding author: ]{zudov@physics.umn.edu}
\affiliation{School of Physics and Astronomy, University of Minnesota, Minneapolis, Minnesota 55455, USA}
\author{L.\,N. Pfeiffer}
\affiliation{Princeton University, Department of Electrical Engineering, Princeton, NJ 08544, USA}
\author{K.\,W. West}
\affiliation{Princeton University, Department of Electrical Engineering, Princeton, NJ 08544, USA}
\received{20 February 2011; published 2 May 2011}

\begin{abstract}
Recent experiments on microwave-irradiated high-mobility two-dimensional electron systems [Dai {\em et al.}, Phys. Rev. Lett. {\bf 105}, 246802 (2010), Hatke {\em et al.}, Phys. Rev B {\bf 83} 121301(R) (2011)] revealed a novel photoresistivity peak in the vicinity of the second cyclotron resonance harmonic.
Here we report on the nonlinear transport measurements and demonstrate that the peak can be {\em induced} by modest dc fields and that its position is {\em not} affected even by strong dc fields, in contrast to microwave-induced resistance oscillations that shift to higher magnetic fields.   
These findings reinforce the notion that the peak cannot be described by existing models and provides important constraints for further theoretical considerations.
\end{abstract} 
\maketitle

Over the past decade, the regime of high Landau levels of high-mobility two-dimensional electron systems (2DESs) revealed a variety of remarkable transport phenomena.
Some prominent examples include microwave-induced resistance oscillations (MIROs),\citep{miro:exp,miro:phase} phonon-induced resistance oscillations,\citep{piro:1} Hall field-induced resistance oscillations,\citep{hiro:1} and several classes of combined oscillations.\citep{comb:1a,comb:1b,khodas:2008}
In addition, very clean microwave- and dc-driven 2DESs can exhibit zero-resistance states\citep{zrs:exp} and zero-differential resistance states,\citep{zdrs:exp} respectively.
These exotic states are believed to originate from instabilities of the 2DES with respect to formation of current domains.\citep{zrs:th}

Mainstream theories of magnetoresistance oscillations are based on quantum kinetics and consider the {\em displacement} mechanism,\citep{miro:th:dsp,hiro:th} originating from the modification of impurity scattering by microwave\citep{miro:th:dsp} or dc\citep{hiro:th} electric fields, and the {\em inelastic} mechanism,\citep{hiro:th,miro:th:in} stepping from the non-equilibrium energy distribution.
Both mechanisms can give rise to MIROs which appear in photoresistivity $\delta\rho_\omega \propto - \Pmw \eac \sin 2\pi\eac$, where
$\Pmw$ is the dimensionless parameter proportional to the microwave power, $\eac\equiv\omega/\oc$, $\omega=2\pi f$ is the microwave frequency, $\oc=eB/m^*$ is the cyclotron frequency, $B$ is the magnetic field, and $m^*$ is the electron effective mass.
Hall field-induced resistance oscillations originate from the oscillatory correction to the differential resistivity due to the displacement mechanism,\citep{hiro:th} $\delta r \propto \cos 2\pi\edc$, where $\edc\equiv e\Edc(2\rc)/\hbar\oc$, $2\rc=2v_F/\wc$ is the cyclotron diameter, $v_F$ is the Fermi velocity, $\Edc=\rho_{H}I/w$ is the Hall electric field, $\rho_H$ is the Hall resistivity, $I$ is the direct current, and $w$ is the sample width.
Under simultaneous application of dc and microwave fields, the resulting oscillations \citep{comb:1a} in differential resistivity are governed by the displacement mechanism and can be described by\citep{khodas:2008}
\begin{equation}
\begin{split}
\label{acdc}
 &\frac{ \delta r_\omega }{ \rho_D }=  \frac{ ( 4 \lambda )^2
\ttr }{\pi \tau_\pi  }\bigg[
( 1 - 2 \Pmw ) \cos 2 \pi \edc
\\
&
+ \Pmw \sum \limits_\pm \pm \frac {(\eac \pm \edc)} \edc \cos 2\pi (\eac \pm \edc) \bigg]\, .
\end{split}
\end{equation}
Here, $\rho_D$ is the Drude resistivity,  $\lambda = \exp(-\pi/\oc\tq)$ is the Dingle factor, and $\tq$, $\ttr$, and $\tpi$ are the quantum, transport, and backscattering lifetimes, respectively.
Equation (\ref{acdc}) well describes recent experiments\citep{comb:1a} showing that the $r_\omega$ maxima occur along the lines defined by $\eac+\edc \simeq n,\,n=1,2,3,...\,$.
This simple relation corresponds to the maximum value of the first term in the sum of \req{acdc} which dominates the response under typical experimental conditions.
In this scenario, the probability of inter-Landau level transitions due to microwave absorption and backscattering off of an impurity in the direction {\em parallel} to the dc field is maximized.  
The functional form of microwave- and Hall field-induced resistance oscillations can be obtained from \req{acdc} by taking the limits $\edc \rightarrow 0$ and $\Pmw \rightarrow 0$, respectively.

Very recently, experiments in irradiated high mobility 2DESs revealed yet another dramatic effect $-$ a distinct photoresistivity peak emerging near the second harmonic of the cyclotron resonance.\citep{dai:2010,hatke:2011b}
In contrast to MIROs, which are known to decay with increasing microwave frequency, this so-called $\xt$ peak appears only above a certain frequency, which is about $100$ GHz in our 2DES.
At higher frequencies, the $\xt$ peak can be more than an order of magnitude stronger than MIROs.\citep{hatke:2011b} 
This remarkable phenomenon {\em cannot} be explained by existing theories \citep{miro:th:dsp,miro:th:in} and its origin remains a mystery.

In this paper, we report on our experimental studies of the nonlinear response of this novel photoresistivity peak in a high-mobility 2DES.
Our measurements are performed at a microwave frequency $f=90$ GHz which is somewhat lower than the critical frequency necessary for the observation of the peak at zero dc fields.
Remarkably, the $\xt$ peak appears in nonlinear differential resistivity under a modest dc field, which apparently helps to separate it from MIROs.
Once appeared, the peak persists over a wide range of dc fields but eventually weakens.
Most importantly, we find that the position of the $\xt$ peak is largely {\em insensitive} to the applied dc field over the whole range of dc fields studied.
This finding is in vast contrast to MIROs that shift substantially to higher magnetic fields as prescribed by \req{acdc}.  
These results strengthen the conclusion that the peak cannot be described by existing theories of microwave photoconductivity and provide important constraints for future theories.

Our sample is a Hall bar ($w = 200$ $\mu$m) fabricated from a symmetrically doped GaAs/Al$_{0.24}$Ga$_{0.76}$As quantum well grown by molecular beam epitaxy.
The density and the mobility were $n_e \simeq 3.4 \times 10^{11}$ cm$^{-2}$ and $\mu \simeq 12.5 \times 10^6$ cm$^2$/Vs.
Microwave radiation was provided by Gunn oscillators feeding frequency doublers.
Measurements were performed in a $^3$He cryostat at a constant coolant temperature $T\simeq 1.5$ K under continuous microwave irradiation in sweeping magnetic fields. 
Differential resistivity $r_\omega$ was measured over a wide range of $I$, from 0 to 100 $\mu$A, using a standard low frequency lock-in technique.

%%%%%%%%%%%%%%%%%%%%%%%%%%%%%%%%%%%%%%%%%%%%%%%%%
\begin{figure}[t]
\includegraphics{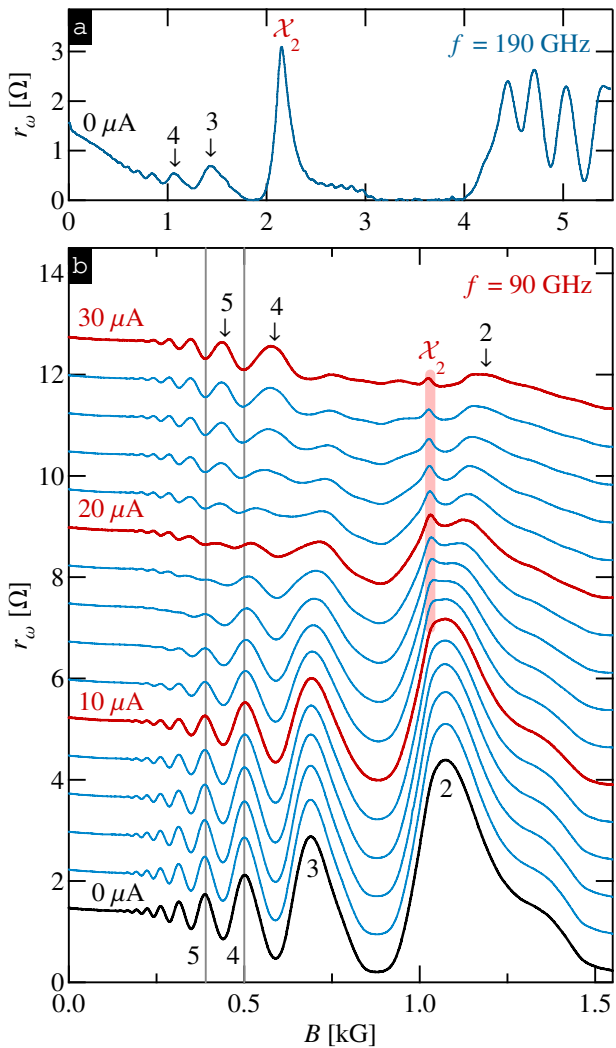}
\caption{(Color online) (a) [(b)] Differential magnetoresistivity $r_\omega(B)$ for direct current $I=0$ $\mu$A [$I=0$ to $I=30$ $\mu$A, bottom to top, with steps of 2 $\mu$A] under microwave irradiation with frequency $f=190$\,[90] GHz.
The traces are vertically offset by 0.75 $\Omega$ and thick lines correspond to a step of 10 $\mu$A (as marked). 
Integers mark the order of the MIRO maxima at $I=0$ and a thick line marks the $\xt$ peak appearing at $I\gtrsim 14$ $\mu$A.
Notice that the MIRO maxima shift to higher $B$ with increasing $I$ (cf. thin vertical lines drawn near $\eac=4,5$).
}
\vsp
\label{fig1}
\end{figure}
%%%%%%%%%%%%%%%%%%%%%%%%%%%%%%%%%%%%%%%%%%%%%%%%%
Before presenting our nonlinear transport data, we first present in  \rfig{fig1}\,(a) the magnetoresistivity measured at $I=0$ and under microwave irradiation of frequency $f=190$ GHz. 
At this high frequency, the data clearly show not only MIROs and associated zero-resistance states, but also a giant $\xt$ peak which is superimposed on the second MIRO maximum.
As mentioned in the introduction, this peak is not observable in our 2DES at $I=0$ when the frequency is below $\simeq 100$ GHz. 
However, as we show next, dc fields make the peak visible even at lower frequencies.

In \rfig{fig1}\,(b) we present the differential magnetoresistivity, $r_\omega(B)$ for direct currents from $I=0$ to $I=30$ $\mu$A (bottom to top, with steps of 2 $\mu$A) measured under microwave irradiation of $f=90$ GHz.
The traces are vertically offset for clarity, and thick lines correspond to a step of 10 $\mu$A.
At zero-bias (bottom trace, $I=0$) the data exhibit pronounced MIROs which persist up to $\eac > 10$. 
The second MIRO maximum, where one might expect to see the $\xt$ peak, does not stand out in any way [cf.\,\rfig{fig1}\,(a)].
However, with increasing current up to $\simeq\,10$ $\mu$A the second MIRO maximum broadens and at higher currents develops a sharp feature (cf.\,thick line).
This feature is very narrow and, as we show below, is located between the second MIRO peak and the second harmonic of the cyclotron resonance. 
Based on its shape and position, we attribute this feature to the $\xt$ peak\citep{dai:2010,hatke:2011b} which appears in our 2DES without a dc bias at higher microwave frequencies [cf.\,\rfig{fig1}\,(a)]. 
The very fact that a modest dc field can cause the appearance of the $\xt$ peak, separating it from MIROs, indicates that its nonlinear response is different from that of MIROs.

Further examination of the data in \rfig{fig1}\,(b) reveals that {\em all} MIRO maxima shift to higher $B$ with increasing current (cf.\,$\downarrow$).
Because most of this shift occurs within a relatively narrow current range, where the oscillation amplitude is strongly suppressed, it can also be viewed as a development of a MIRO maximum into a minimum at a given $\eac$.
Indeed, the fourth and fifth MIRO maxima at $I=0$ both evolve into the minima at $I=30$ $\mu$A (cf.\,thin vertical lines).
This evolution with increasing $I$ is consistent with earlier experiments\citep{comb:1a} and is well understood within the displacement model\citep{khodas:2008} [cf. \req{acdc}].
In light of such dramatic changes to the MIRO waveform, it is quite remarkable to see that the $\xt$ peak, once developed, {\em does not} change either its position or width in this range of currents (cf.\,thick line). 
As we show below, even higher dc fields do not alter the position of the $\xt$ peak.

%%%%%%%%%%%%%%%%%%%%%%%%%%%%%%%%%%%%%%%%%%%%%%%%%
\begin{figure}[t]
\includegraphics{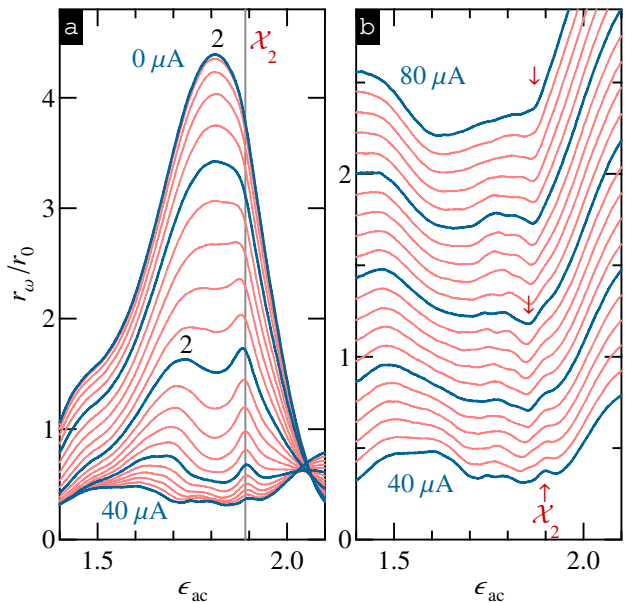}
\caption{(Color online) (a)[(b)] Normalized differential resistivity $r_\omega(\eac)/r_0$ for direct currents from $I=0$ $\mu$A to $I=40$ $\mu$A [$I=40$ $\mu$A to $I=80$ $\mu$A, with steps of 2 $\mu$A] under microwave irradiation of $f=90$ GHz.
In panel (b), the traces are vertically offset by 0.1 for clarity.
Thick traces correspond to a step of 10 $\mu$A.
Notice that the position of the $\xt$ peak is independent of $I$ [cf. vertical line in (a)], while the second MIRO peak shifts to lower $\eac$.
}
\vsp
\label{fig2}
\end{figure}
%%%%%%%%%%%%%%%%%%%%%%%%%%%%%%%%%%%%%%%%%%%%%%%%%
In \rfig{fig2}\,(a) we replot the differential resistivity, now normalized to its value at $B=0$, $r_\omega/r_0$, as a function of $\eac$ for direct currents from $I=0$ to $I=40$ $\mu$A in steps of 2 $\mu$A under microwave irradiation of $f=90$ GHz.
Plotted in such a way, the data clearly demonstrate that the $\xt$ peak occurs at $\eac=\eac^{\xt}$ (cf.\,vertical line), which falls {\em in between} the second MIRO maximum and the second harmonic of the cyclotron resonance, $\eac^{2+} < \eac^{\xt} < 2$.
The data further show that in the range between 30 $\mu$A and 40 $\mu$A the magnitude of the $\xt$ peak is reduced considerably but its position remains unchanged.

In \rfig{fig2}\,(b) we present the results obtained at still higher currents, from $I=40$ $\mu$A (bottom curve) to $I=80$ $\mu$A (top curve) in steps of 2 $\mu$A.
The data are vertically offset for clarity by 0.1, and thick lines again correspond to a step of 10 $\mu$A.
While the $\xt$ peak continues to decay with increasing current, it remains clearly visible up to $I\simeq 60$ $\mu$A.
Concurrent with the decay of the $\xt$ peak, a rather sharp minimum develops at a slightly lower $\eac$ (cf.\,$\downarrow$).
At higher currents, $I\gtrsim 60 $ $\mu$A, this minimum becomes the most pronounced feature in close vicinity to the original $\xt$ peak (cf.\,$\downarrow$ at the top trace).
Based on the conversion of the MIRO maxima into the minima, as illustrated in \rfig{fig1}\,(b), one might consider a possibility that this minimum originates from the $\xt$ peak in a similar way but with a very different energy scale involved.

\begin{figure}[b]
\includegraphics{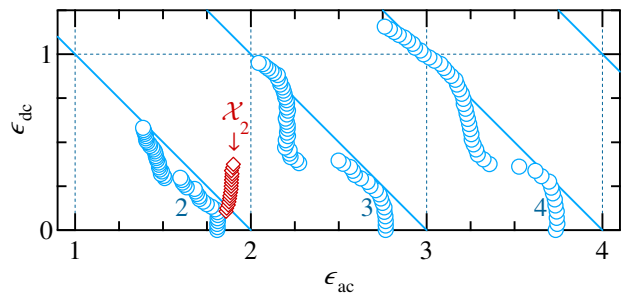}
\vspace{-0.1 in}
\caption{(Color online) Positions of the maxima of differential resistivity $r_\omega$ corresponding to MIROs, $\eac^{2+},\eac^{3+},\eac^{4+}$ (cf.\,$2,3,4$), and to the $\xt$ peak (cf.\,$\xt$) on a ($\eac,\edc$) plane.
Solid lines are drawn at $\eac+\edc=n,\, n=2,3,4,5$.  
}
\vsp
\label{fig3}
\end{figure}

To summarize our observations, we now extract both $\eac$ and $\edc$ from our experimental data for the selected $r_\omega(B)$ maxima obtained at direct currents up to 100 $\mu$A.
Figure \ref{fig3} shows the results of this extraction on a ($\eac,\edc$) plane both for the MIRO maxima (cf.\,$2,3,4$) and for the $\xt$ peak (cf.\,$\xt$). 
At small $\edc$, MIRO maxima appear\citep{miro:phase} at $\eac \simeq n-\varphi_n$, where $\varphi_2\simeq 0.19$ and $\varphi_3 \simeq \varphi_4 \simeq 0.25$.
At somewhat higher dc fields, these maxima are well described by a linear dependence, $\eac+\edc=n$ (cf.\,solid lines), in agreement with \req{acdc}.
At $\edc\simeq 1/2$, roughly corresponding to the first minimum of Hall field-induced resistance oscillations, higher order ($n =3,4$) MIRO maxima quickly jump to $\eac \simeq (n-1)+1/4$, a position for the MIRO minima which remains satisfied for $1/2\lesssim \edc \lesssim 1-1/4$.
This jump is a result of the maxima (minima) conversion into the minima (maxima) as seen in \rfig{fig1}\,(b).
At still higher $\edc$, maxima again follow linear dependence, $\eac+\edc=n$.
 
In contrast to MIROs, the $\xt$ peak exhibits very different behavior.
Once developed at $\edc\simeq 0.1$, the $\xt$ peak is found at $\eac=2-\varphi$, where $\varphi < \varphi_2$. 
At higher dc fields, the position of the $\xt$ peak remains essentially unchanged ($\eac^{\xt}\simeq {\rm const}$) over the whole range of currents over which the peak is observed.
A closer examination of the data reveals that the $\xt$ peak, in fact, moves slightly towards the second harmonic of the cyclotron resonance, the direction which is {\em opposite} to that of MIROs.
However, this move can be caused by a fast decay of the neighboring MIRO peak and by its fast movement towards lower $\eac$ with increasing dc field.

%%%%%%%%%%%%%%%%%%%%%%%%%%%%%%%%%%%%%%%%%%%%%%%%%
%%%%%%%%%%%%%%%%%%%%%%%%%%%%%%%%%%%%%%%%%%%%%%%%%
In summary, we have studied the nonlinear response of irradiated high-mobility 2DESs, focusing on the recently discovered\citep{dai:2010,hatke:2011b} photoresistivity peak, which appears in the vicinity to the second harmonic of the cyclotron resonance and high microwave frequencies.
We have found that at microwave frequencies slightly below the minimum frequency necessary for the observation of this peak at zero dc field, the peak appears in nonlinear differential resistivity under a modest dc field, which apparently helps to separate it from MIROs.
Once developed, the peak persists over a wide range of dc fields and eventually disappears.
Most importantly, the position of this $\xt$ peak is {\em not} changed significantly over the whole range of dc fields.
This behavior is in contrast to the evolution of microwave-induced resistance oscillations that shift to higher magnetic fields in accordance with the displacement model, as prescribed by \req{acdc}.   
Our findings further indicate that the nature of the $\xt$ peak is different from that of MIROs and provide important constraints for theoretical considerations.

We thank I. Dmitriev, M. Khodas, and B. Shklovskii for discussions, and S. Hannas, G. Jones, J. Krzystek, T. Murphy, E. Palm, J. Park, D. Smirnov, and A. Ozarowski for technical assistance.
A portion of this work was performed at the National High Magnetic Field Laboratory, which is supported by NSF Cooperative Agreement No. DMR-0654118, by the State of Florida, and by the U.S. Department of Energy (DOE).
The work at Minnesota was supported by the DOE Grant No. DE-SC0002567 (high frequency measurements at NHMFL) and by the NSF Grant No. DMR-0548014 (low frequency measurements at Minnesota). 
The work at Princeton was partially funded by the Gordon and Betty Moore Foundation as well as the NSF MRSEC Program through the Princeton Center for Complex Materials (DMR-0819860).
A.T.H. acknowledges support from the University of Minnesota.

%\newpage

\end{document}